\documentclass[titlepage]{article}%
\usepackage{amsmath}
\usepackage{cite}
\usepackage{graphicx}
\usepackage{color}%
\usepackage{amsfonts}%
\usepackage{amssymb}
\newtheorem{theorem}{Theorem}

\newtheorem{lemma}[theorem]{Lemma}

\newenvironment{proof}[1][Proof]{\noindent\textbf{#1.} }{\ \rule{0.5em}{0.5em}}
\begin{document}

\title{Generation interval contraction and epidemic data analysis}
\author{Eben Kenah$^{1,2,\ast}$, Marc Lipsitch$^{1,3}$, James M. Robins$^{1,2}$\\$^{1}$Department of Epidemiology\\$^{2}$Department of Biostatistics\\$^{3}$Department of Immunology and Infectious Disease\\Harvard School of Public Health\\677 Huntington Ave., Boston, Massachusetts, USA\\*Corresponding author: ekenah@hsph.harvard.edu}
\date{April and November, 2006\\
Revised January-June and October-November, 2007}
\maketitle

\begin{abstract}
The \textit{generation interval} is the time between the infection time of an
infected person and the infection time of his or her infector. \ Probability
density functions for generation intervals have been an important input for
epidemic models and epidemic data analysis. \ In this paper, we specify a
general stochastic SIR epidemic model and prove that the mean generation
interval decreases when susceptible persons are at risk of infectious contact
from multiple sources. \ The intuition behind this is that when a susceptible
person has multiple potential infectors, there is a \textquotedblleft
race\textquotedblright\ to infect him or her in which only the first
infectious contact leads to infection. \ In an epidemic, the mean generation
interval contracts as the prevalence of infection increases. \ We call this
\textit{global competition} among potential infectors. \ When there is rapid
transmission within clusters of contacts, generation interval contraction can
be caused by a high local prevalence of infection even when the global
prevalence is low. \ We call this \textit{local\ competition} among potential
infectors. \ Using simulations, we illustrate both types of competition.
\ Finally, we show that hazards of infectious contact can be used instead of
generation intervals to estimate the time course of the effective reproductive
number in an epidemic. \ This approach leads naturally to partial likelihoods
for epidemic data that are very similar to those that arise in survival
analysis, opening a promising avenue of methodological research in infectious
disease epidemiology. \ \ 

\end{abstract}

\section{Introduction}

In infectious disease epidemiology, the \textit{serial interval} is the
difference between the symptom onset time of an infected person and the
symptom onset time of his or her infector \cite{Giesecke}. \ This is sometimes
called the \textquotedblleft generation interval.\textquotedblright%
\ \ However, we find it more useful to adopt the terminology of Svensson
\cite{Svensson} and define the \textit{generation interval }as the difference
between the infection time of an infected person and the infection time of his
or her infector. \ By these definitions, the serial interval is observable
while the generation interval usually is not. \ We define \textit{infectious
contact} from $i$ to $j$ to be a contact that is sufficient to infect $j$ if
$i$ is infectious and $j$ is susceptible, and we define a \textit{potential
infector} of person $i$ to be an infectious person who has positive
probability of making infectious contact with $i$. \ Finally, we use the term
\textit{hazard} rather than \textit{force of infection} to highlight the
similarities between epidemic data analysis and survival analysis.

The generation interval has been an important input for epidemic models used
to investigate the transmission and control of SARS \cite{Lipsitch,Wallinga}
and pandemic influenza \cite{Mills,Ferguson}. \ More recently, generation
interval distributions have been used to calculate the incubation period
distribution of SARS \cite{KukMa} and to estimate $R_{0}$ from the exponential
growth rate at the beginning of an epidemic \cite{Wallinga2}. \ It is
generally assumed that the generation interval distribution is characteristic
of an infectious disease. \ In this paper, we show that this is not true.
\ Instead, the expected generation interval decreases as the number of
potential infectors of susceptibles increases. \ During an epidemic,
generation intervals tend to contract as the prevalence of infection
increases. \ This effect was described by Svensson \cite{Svensson} for an SIR
model with homogeneous mixing. \ In this paper, we extend this result to all
time-homogeneous stochastic SIR\ models.

A simple thought experiment illustrates the intuition behind our main result.
\ Imagine a susceptible person $j$ in a room. \ Place $m$ other persons in the
room and infect them all at time $t=0$. \ For simplicity, assume that
infectious contact from $i$ to $j$ occurs with probability one, $i=1,...,m$.
\ Let $t_{ij}$ be a continuous nonnegative random variable denoting the first
time at which $i$ makes infectious contact with $j$. \ Person $j$ is infected
at time $t_{j}=\min(t_{1j},...,t_{mj})$. \ Since all infectious persons were
infected at time zero, $t_{j}$ is the generation interval. \ If we repeat the
experiment with larger and larger $m$, the expected value of $\min
(t_{1j},...,t_{mj})$ will decrease.

When a susceptible person is at risk of infectious contact from multiple
sources, there is a \textquotedblleft race\textquotedblright\ to infect him or
her in which only the first infectious contact leads to infection.
\ Generation interval contraction is an example of a well-known phenomenon in
epidemiology: The expected time to an outcome, given that the outcome occurs,
decreases in the presence of competing risks. \ In our thought experiment, the
outcome is the infection of $j$ by a given $i$ and the competing risks are
infectious contacts from all sources other than $i$. \ 

Adapting our thought experiment slightly, we see that the contraction of the
generation interval is a consequence of the fact that the hazard of infection
for $j$ increases as the number of potential infectors increases. \ Let
$\lambda(t)$ be the hazard of infectious contact from any potential infector
to $j$ at time $t$ and let $E[t_{j}|m]$ be the expected infection time of $j$
given $m$ potential infectors. \ Then
\begin{align*}
E[t_{j}|m]  & =\int_{0}^{\infty}e^{-m\lambda(t)}dt\\
& >\int_{0}^{\infty}e^{-(m+1)\lambda(t)}dt=E[t_{j}|m+1],
\end{align*}
so the expected generation interval decreases as the number of potential
infectors increases. \ A hazard of infection that increases with the number of
potential infectors is a defining feature of most epidemic models, so
generation interval contraction is a very general phenomenon. \ We note that a
very similar phenomenon occurs in endemic diseases, where increased force of
infection results in a decreased average age at first infection
\cite{AndersonMay}.

The rest of the paper is organized as follows: In Section 2, we describe a
general stochastic SIR epidemic model. \ In Section 3, we use this model to
show that the mean generation interval decreases as the number of potential
infectors increases. \ As a corollary, we find that the mean serial interval
also decreases. \ In Section 4, we consider the role of the population contact
structure in generation interval contraction and illustrate the effects of
global and local competition among potential infectors with simulations. \ In
Section 5, we argue that hazards of infectious contact should be used instead
of generation or serial interval distributions in the analysis of epidemic
data. \ Section 6 summarizes our main results and conclusions.

\section{General stochastic SIR model}

We start with a very general stochastic "Susceptible-Infectious-Removed" (SIR)
epidemic model. \ This model includes fully-mixed and network-based models as
special cases, and it has been used previously to define a mapping from the
final outcomes of stochastic SIR\ models to the components of semi-directed
random networks \cite{Kenah1,Kenah2}.

Each person $i$ is infected at his or her \textit{infection time} $t_{i}$,
with $t_{i}=\infty$ if $i$ is never infected. \ Person $i$ recovers from
infectiousness or dies at time $t_{i}+r_{i}$, where the \textit{recovery
period} $r_{i}$ is a positive random variable with the cumulative distribution
function (cdf) $F_{i}(r)$. \ The recovery period $r_{i}$ may be the sum of a
\textit{latent period}, during which $i$ is infected but not infectious, and
an \textit{infectious period}, during which $i$ can transmit infection. \ We
assume that all infected persons have a finite recovery period. \ If person
$i$ is never infected, let $r_{i}=\infty$. \ Let Sus$(t)=\{i:t_{i}>t\}$ be the
set of susceptibles at time $t$. \ 

When person $i$ is infected, he or she makes infectious contact with person
$j$ after an \textit{infectious contact interval} $\tau_{ij}$. \ Each
$\tau_{ij}$ is a positive random variable with cdf $F_{ij}(\tau|r_{i})$ and
survival function $S_{ij}(\tau|r_{i})=1-F_{ij}(\tau|r_{i})$. \ Let $\tau
_{ij}=\infty$ if person $i$ never makes infectious contact with person $j$, so
the infectious contact interval distribution may have probability mass at
$\infty$. \ Define
\[
S_{ij}(\infty|r_{i})=\lim_{\tau\rightarrow\infty}S_{ij}(\tau|r_{i}),
\]
which is the conditional probability that $i$ never makes infectious contact
with $j$ given $r_{i}$. \ Since a person cannot transmit disease before being
infected or after recovering from infectiousness, $S_{ij}(\tau|r_{i})=1$ for
all $\tau\leq0$ and $S_{ij}(\tau|r_{i})=S_{ij}(\infty|r_{i})$ for all
$\tau\geq r_{i}$. \ Since a person cannot infect himself (or herself),
$\tau_{ii}=\infty$ with probability one and $S_{ii}(\tau|r_{i})=1$ for all
$\tau$. \ 

\begin{figure}
[ptb]
\begin{center}
\includegraphics[
height=1.2064in,
width=5.5486in
]%
{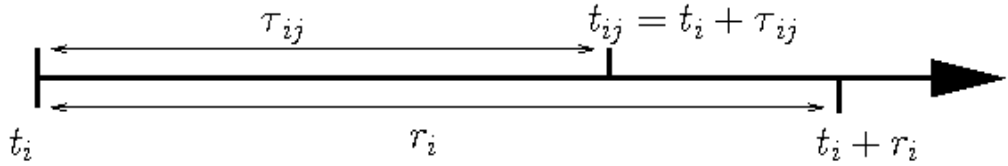}%
\caption{Schematic diagram of variables in the general stochastic SIR model
for the ordered pair $ij$. Recall that $t_{j}\leq t_{ij}$. As discussed in
Section 3.2, person $i$ develops symptoms at time $t_{i}^{\mathrm{sym}}%
=t_{i}+q_{i}$, where $q_{i}$ is the incubation period.}%
\label{schematic}%
\end{center}
\end{figure}

The \textit{infectious contact time} $t_{ij}=t_{i}+\tau_{ij}$ is the time at
which person $i$ makes infectious contact with person $j$. \ If person $j$ is
susceptible at time $t_{ij}$, then $i$ infects $j$ and $t_{j}=t_{ij}$. \ If
$t_{ij}<\infty$, then $t_{j}\leq t_{ij}$ because person $j$ avoids infection
at time $t_{ij}$ only if he or she has already been infected. \ If person $i$
never makes infectious contact with person $j$, then $t_{ij}=\infty$ because
$\tau_{ij}=\infty$. \ Figure \ref{schematic} shows a schematic diagram of the
relationships among $r_{i}$, $\tau_{ij}$, and $t_{ij}$. \ 

The \textit{importation time} $t_{0i}$ of person $i$ is the earliest time at
which he or she receives infectious contact from outside the population. \ The
importation time vector $\mathbf{t}_{0}=(t_{01},...,t_{0n})$. \ 

We assume that each infected person has a unique infector. \ Following
\cite{Wallinga}, we let $v_{i}$ represent the index of the person who infected
person $i$, with $v_{i}=0$ for imported infections and $v_{i}=\infty$ if $i$
is never infected. \ If tied infectious contact times have nonzero
probability, then $v_{i}$ can be chosen from all $j$ such that $t_{ji}%
=t_{i}<\infty$. \ 

\subsection{Epidemics}

Let $t_{(1)}\leq t_{(2)}\leq...\leq t_{(m)}$ be the order statistics of all
$t_{1},...,t_{n}$ less than infinity, and let $(k)$ be the index of the
$k^{\text{th}}$ person infected. \ Before the epidemic begins, an importation
time vector $\mathbf{t}_{0}$ is chosen. \ The epidemic begins at time
$t_{(1)}=\min_{i}(t_{0i})$. \ Person $(1)$ is assigned a recovery time
$r_{(1)}$. \ Every person $j\in\,$Sus$(t_{(1)})$ is assigned an infectious
contact time $t_{(1)j}=t_{(1)}+\tau_{(1)j}$. \ The second infection occurs at
$t_{(2)}=\min_{j\in\text{Sus}(t_{(1)})}\min(t_{0j},t_{(1)j})$, which is the
first infectious contact time after $t_{(1)}$. \ Person $(2)$ is assigned a
infectious period $r_{(2)}$. \ After $k$ infections, the next infection occurs
at $t_{(k+1)}=\min_{j\in\text{Sus}(t_{(k)})}\min(t_{0j},t_{(1)j}%
,...,t_{(k)j})$. \ The epidemic stops after $m$ infections if and only if
$t_{(m+1)}=\infty$. \ 

\section{Generation interval contraction}

In this section, we show that the mean infectious contact interval $\tau_{ij}$
given that $i$ infects $j$ is shorter than the mean infectious contact
interval given that $i$ makes infectious contact with $j$. \ In the notation
from the previous section,
\[
E[\tau_{ij}|v_{j}=i]\leq E[\tau_{ij}|\tau_{ij}<\infty]
\]
(note that $v_{j}=i$ implies $\tau_{ij}<\infty$ but not vice versa). \ In
general, this inequality is strict when $j$ is at risk of infectious contact
from any source other than $i$. \ This inequality implies the contraction of
generation and serial intervals during an epidemic. \ For background on the
probability theory used in this section, please see Ref. \cite{Gut} or any
other probability text.

\begin{lemma}
$E[\tau_{ij}|v_{j}=i]\leq E[\tau_{ij}|\tau_{ij}<\infty].$
\end{lemma}

\begin{proof}
We first show that $E[\tau_{ij}|r_{i},\tau_{ij}<\infty]\leq E[\tau_{ij}%
|r_{i},v_{j}=i]$ and then use the law of iterated expectation. \ If person $i$
was infected at time $t_{i}$ and has recovery period $r_{i}$, then the
probability that $\tau_{ij}<\infty$ is $F_{ij}(\infty|r_{i})=1-S_{ij}%
(\infty|r_{i})$. \ Let
\[
F_{ij}^{\ast}(\tau|r_{i})=\frac{F_{ij}(\tau|r_{i})}{F_{ij}(\infty|r_{i})}%
\]
be the conditional cdf of $\tau_{ij}$ given $r_{i}$ and $\tau_{ij}<\infty$.
\ Then
\begin{equation}
E[\tau_{ij}|r_{i},\tau_{ij}<\infty]=\int_{0}^{r_{i}}\tau dF_{ij}^{\ast}%
(\tau|r_{i}).\label{Etau}%
\end{equation}
If person $j$ is susceptible at time $t_{i}$ and $\tau_{ij}<\infty$, then
$v_{j}=i$ if and only if $j$ escapes infectious contact from all other
infectious people during the time interval $(t_{i},t_{i}+\tau_{ij})$. \ Let
$S_{\ast j}(t_{i}+\tau)$ be the probability that $j$ escapes infectious
contact from all sources other than $i$ in the interval $(t_{i},t_{i}+\tau)$.
\ Given $r_{i}$ and $\tau_{ij}<\infty$, the conditional probability density
for an infectious contact from $i$ to $j$ at time $t_{i}+\tau$ that leads to
the infection of $j$ is proportional to
\[
S_{\ast j}(t_{i}+\tau)dF_{ij}^{\ast}(\tau|r_{i}).
\]
If we let
\[
\psi=\int_{0}^{r_{i}}S_{\ast j}(t_{i}+\tau)dF_{ij}^{\ast}(\tau|r_{i}),
\]
then
\[
E[\tau_{ij}|r_{i},v_{j}=i]=\int_{0}^{r_{i}}\tau\frac{S_{\ast j}(t_{i}+\tau
)}{\psi}dF_{ij}^{\ast}(\tau|r_{i}).
\]
Since $S_{\ast j}(t_{i}+\tau_{ij})$ is a monotonically decreasing function of
$\tau_{ij}$,
\begin{align*}
E[\tau_{ij}|r_{i},v_{j}=i]-E[\tau_{ij}|r_{i},\tau_{ij}<\infty]=~ &
E[\tau_{ij}\frac{S_{\ast j}(t_{i}+\tau_{ij})}{\psi}|r_{i},\tau_{ij}<\infty]\\
&  -E[\tau_{ij}|r_{i},\tau_{ij}<\infty]E[\frac{S_{\ast j}(t_{i}+\tau_{ij}%
)}{\psi}|r_{i},\tau_{ij}<\infty]\\
=~ &  \text{Cov}(\tau_{ij},\frac{S_{\ast j}(t_{i}+\tau_{ij})}{\psi}|r_{i}%
,\tau_{ij}<\infty)\\
\leq &  ~0.
\end{align*}
Therefore,
\begin{equation}
E[\tau_{ij}|r_{i},v_{j}=i]\leq E[\tau_{ij}|r_{i},\tau_{ij}<\infty
].\label{ineq0}%
\end{equation}
Since the same inequality holds for all $r_{i}$,
\begin{align}
E[\tau_{ij}|v_{j}=i]  & =E[E[\tau_{ij}|r_{i},v_{j}=i]]\nonumber\\
& \leq E[E[\tau_{ij}|r_{i},\tau_{ij}<\infty]]=E[\tau_{ij}|\tau_{ij}%
<\infty]\label{Ineq}%
\end{align}
by the law of iterated expectation. \ 
\end{proof}

Equality holds in equation (\ref{ineq0}) if and only if $\tau_{ij}$ and
$S_{\ast j}(t_{i}+\tau_{ij})$ have covariance zero given $r_{i}$ and
$\tau_{ij}<\infty$. \ Since $S_{\ast j}(t_{i}+\tau_{ij})$ is a monotonically
decreasing function of $\tau_{ij}$, this will occur if and only if $\tau_{ij}$
or $S_{\ast j}(t_{i}+\tau_{ij})$ is constant given $r_{i}$ and $\tau
_{ij}<\infty$. \ Equality holds in equation (\ref{Ineq}) if and only if
equality holds in (\ref{ineq0}) with probability one in $r_{i}$. $\ $If
$\tau_{ij}$ is constant, then clearly $S_{\ast j}(t_{i}+\tau_{ij})$ is
constant and their covariance is zero. \ If $j$ is not at risk of infectious
contact from any source other than $i$, then $S_{\ast j}(t_{i}+\tau_{ij})$
will be constant even when $\tau_{ij}$ is not. \ In the thought experiment
from the Introduction, the expected infection time of the susceptible $j$
would remain constant in the following two scenarios: (i) all infectious
persons make infectious contact with $j$ at a fixed time $t_{0}$, or \ (ii)
$j$ is only at risk of infectious contact from a single person. \ Scenario (i)
corresponds to a constant $\tau_{ij}$ and scenario (ii) corresponds to a
constant $S_{\ast j}(t_{i}+\tau_{ij})$.

The expected generation interval from $i$ to $j$ given $v_{j}=i$ will be
shortest when the risk of infectious contact to $j$ from sources other than
$i$ is greatest. \ More specifically,
\[
E[\tau_{ij}|r_{i},v_{j}=i]-E[\tau_{ij}|r_{i},\tau_{ij}<\infty]
\]
will be minimized when $S_{\ast j}(t_{i}+\tau_{ij})$ decreases fastest in
$\tau_{ij}$. \ In general, the risk of infectious contact from other sources
will be greatest when the prevalence of infection is highest, so we expect the
greatest contraction of the serial interval during an epidemic to coincide
with the peak prevalence of infection.

In general, we expect to see the following pattern over the course of an
epidemic: The mean generation interval decreases as the prevalence of
infection increases, reaches a minimum as the prevalence of infection peaks,
and increases again as the prevalence of infection decreases. \ 

\subsection{Types of generation intervals}

In \cite{Svensson}, Svennson discussed two types of generation intervals that
are consistent with the verbal definition given in the Introduction. \ $T_{p}$
($p$ for \textquotedblleft primary\textquotedblright) denotes $\tau_{ij}$
where $i$ is chosen at random from all persons who infect at least one other
person and $j$ is chosen randomly from the set of persons $i$ infects.
\ $T_{s}$ ($s$ for \textquotedblleft secondary\textquotedblright) denotes
$\tau_{ij}$ where $j$ is chosen at random from all persons infected from
within the population and $i=v_{j}$. \ $T_{p}$ and $T_{s}$ differ only in the
sampling procedure used to obtain the ordered pair $ij$; $T_{p}$ samples
primary cases (infectors) at random while $T_{s}$ samples secondary cases at
random. \ Equation (\ref{Ineq}) implies that both $E[T_{p}]$ and $E[T_{s}]$
decrease when susceptible persons are at risk of infectious contact from
multiple sources. \ This contraction occurs because the definitions of $T_{p}$
and $T_{s}$ include only $\tau_{ij}$ such that $i$ actually infected $j$. \ 

\subsection{Serial interval contraction}

In an epidemic, infection times are generally unobserved. \ Instead, symptom
onset times are observed. \ Recall that the time between the onset of symptoms
in an infected person and the onset of symptoms in his or her infector is
called the \textit{serial interval}. \ Contraction of the mean generation
interval implies contraction of the mean serial interval as well. \ The
\textit{incubation period} is the time from infection to the onset of symptoms
\cite{Giesecke}. \ Let $q_{i}$ be the incubation period in person $i $, and
let $t_{i}^{\text{sym}}=t_{i}+q_{i}$ be the time of his or her onset of
symptoms. \ If $v_{j}=i$, then the serial interval associated with person $j$
is
\[
t_{j}^{\text{sym}}-t_{i}^{\text{sym}}=\tau_{ij}+q_{j}-q_{i}.
\]
Therefore,
\begin{align*}
E[t_{j}^{\text{sym}}-t_{i}^{\text{sym}}|v_{j}=i]  & =E[\tau_{ij}%
|v_{j}=i]+E[q_{j}]-E[q_{i}]\\
& \leq E[\tau_{ij}|\tau_{ij}<\infty]+E[q_{j}]-E[q_{i}],
\end{align*}
with strict inequality whenever strict inequality holds for the corresponding
generation interval. \ Over the course of an epidemic, we expect the mean
serial interval to follow a pattern very similar to that of the mean
generation interval.

\section{Simulations}

We refer to the \textquotedblleft race\textquotedblright\ to infect a
susceptible person as \textit{competition among potential infectors}. \ In
this section, we illustrate two types of competition among potential
infectors: \textit{Global competition} among potential infectors results from
a high global prevalence of infection. \ \textit{Local competition} among
potential infectors results from rapid transmission within clusters of
contacts, which causes susceptibles to be at risk of infectious contact from
multiple sources within their clusters even if the global prevalence of
infection is low. \ In real epidemics, the prevalence of infection is usually
low but there is clustering of contacts within households, hospital wards,
schools, and other settings. \ 

In this section, we use simulations to illustrate generation interval
contraction under global and local competition among potential infectors.
\ Each simulation is a single realization of a stochastic SIR model in a
population of $10,000$. \ We keep track of the infection times of the primary
and secondary case in each infector/infectee pair and the prevalence of
infection at the infection time of the secondary case, which is a proxy for
the amount of competition to infect the secondary case. \ We then calculate a
smoothed mean of the generation interval as a function of the infection time
of the primary case in each pair. \ Another valid approach would be to
calculate the smoothed means from the results of many simulations. \ We did
not take this approach for the following reasons: (i) Because of variation in
the time course of different realizations of the same stochastic SIR\ model,
many simulations would be required to obtain a curve that reliably
approximates the asymptotic limit. (ii) The smoothed mean over many
simulations would show a pattern similar to that obtained in any single
simulation. \ (iii) Generation interval contraction was proven in Section 3,
so the simulations are intended primarily as illustrations.

All simulations were implemented in Mathematica 5.0.0.0 [\copyright 1988-2003
Wolfram Research, Inc.]. \ All data analysis was done using Intercooled Stata
9.2 [\copyright \ 1985-2007 StataCorp LP] \ All smoothed means are running
means with a bandwidth of $0.8$ (the default for the Stata command
\texttt{lowess} with the option \texttt{mean}). \ Similar results were
obtained for larger and smaller bandwidths.

\subsection{Global competition}

To illustrate global competition among potential infectors, we use a
fully-mixed model with population size $n=10,000$ and basic reproductive
number $R_{0}$. \ The infectious period is fixed, with $r_{i}=1$ with
probability one for all $i$. \ The infectious contact intervals $\tau_{ij}$
have an exponential distribution with hazard $R_{0}(n-1)^{-1}$ truncated at
$r_{i}$, so $S_{ij}(\tau|r_{i})=e^{-R_{0}(n-1)^{-1}\tau}$ when $0<\tau<1$ and
$\tau_{ij}=\infty$ with probability $e^{-R_{0}(n-1)^{-1}}$. \ The epidemic
starts with a single imported infection and no other imported infections
occur. \ 

From equation (\ref{Etau}), the mean infectious contact interval given that
contact occurs is
\[
E[\tau_{ij}|\tau_{ij}<\infty]=\int_{0}^{1}\frac{e^{-R_{0}\tau(n-1)^{-1}%
}-e^{-R_{0}(n-1)^{-1}}}{1-e^{-R_{0}(n-1)^{-1}}}d\tau
\]
For $n=10,000$, Table \ref{Expected} shows this expected value at each $R_{0}
$. \ For all $R_{0}$, $E[\tau_{ij}|\tau_{ij}<\infty]\approx.5$. \ 

\begin{figure}
[ptb]
\begin{center}
\includegraphics[
height=3.7602in,
width=5.5478in
]%
{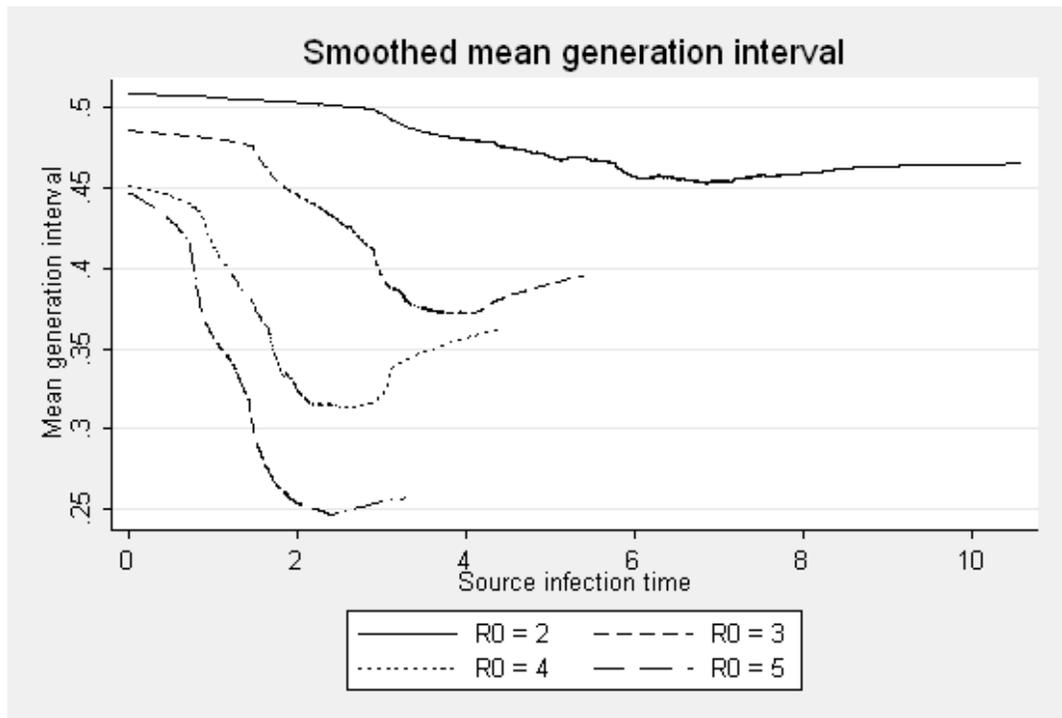}%
\caption{The smoothed mean generation interval as a function the source
infection time for $R_{0}=2,3,4,5$. \ There is a clear tendency to contract,
with greater contraction for higher $R_{0}$. }%
\label{serialintervals2345}%
\end{center}
\end{figure}

\begin{figure}
[ptb]
\begin{center}
\includegraphics[
height=3.7602in,
width=5.5478in
]%
{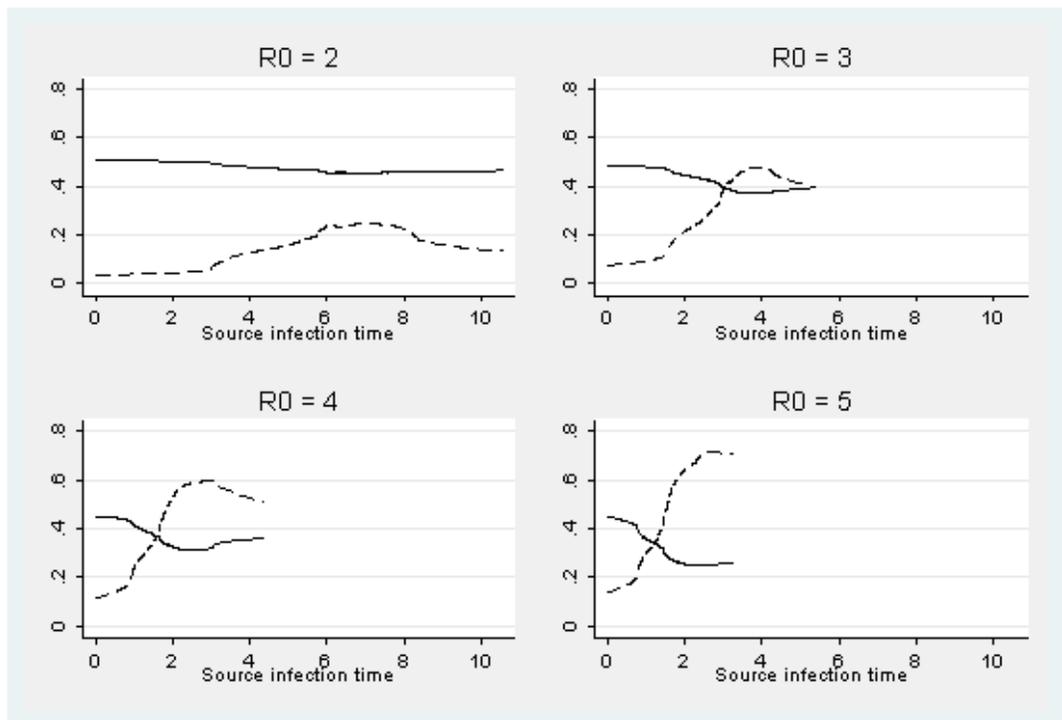}%
\caption{The smoothed mean generation interval (solid lines) and prevalence
(dotted lines) as a function of the source infection time for $R_{0}=2,3,4,5$.
\ In all cases, the greatest contraction of the serial interval coincides with
the peak prevalence of infection (i.e., the greatest competition among
potential infectors).}%
\label{prevalence2345}%
\end{center}
\end{figure}

\begin{figure}
[ptb]
\begin{center}
\includegraphics[
height=3.7602in,
width=5.5478in
]%
{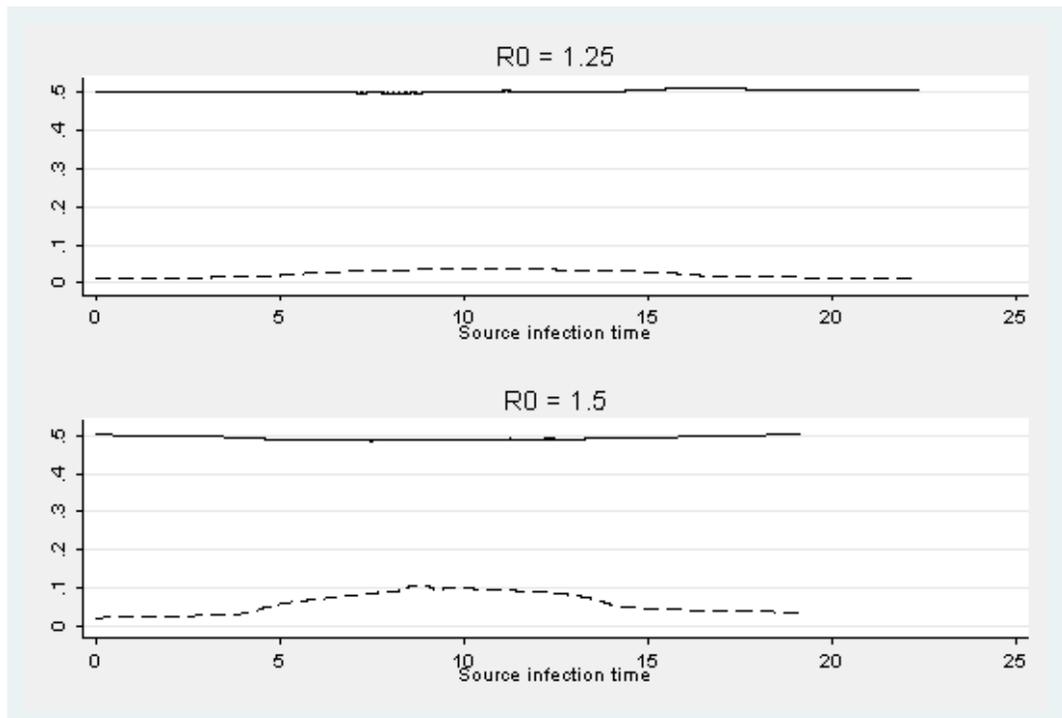}%
\caption{The smoothed mean generation intervals (solid lines) and prevalence
(dotted lines) as a function of the source infection time for $R_{0}=1.25$ and
$1.50$. \ For $R_{0}$ near one, the mean generation interval stays relatively
constant. }%
\label{prevalence1}%
\end{center}
\end{figure}

This model was run once at $R_{0}=1.25$, $1.5$, $2$, $3$, $4$, $5$, and $10$.
\ For each simulation, we recorded $t_{i}$, $v_{i}$, $t_{v_{i}}$, and the
prevalence of infection at time $t_{i}$ in each infector/infectee pair.
\ Figure \ref{serialintervals2345} shows smoothed mean curves for the
generation interval versus the source infection time for $R_{0}=2,3,4,5$.
\ There is a clear tendency for the mean generation interval to contract, with
greater contraction at higher $R_{0}$. \ Figure \ref{prevalence2345} shows
smoothed mean curves for the generation interval and the prevalence of
infection versus the source infection time at each $R_{0}$; in each case, the
greatest contraction of the serial interval coincides with the peak prevalence
of infection (i.e., the greatest competition among potential infectors).
\ Figure \ref{prevalence1} shows the same curves for $R_{0}=1.25$ and $1.50$;
in these cases, the generation interval stays relatively constant. \ These
results are exactly in line with the argument of Section 3. \ 

\subsection{Local competition}

To illustrate local competition among potential infectors, we grouped a
population of $n=9,000$ individuals into clusters of size $k$. \ As before,
the infectious period is fixed at $r_{i}=1$ for all $i$. \ When $i$ and $j$
are in the same cluster, the infectious contact interval $\tau_{ij}$ has an
exponential distribution with hazard $\lambda_{\text{within}}$ truncated at
$r_{i}$, so $S_{ij}(\tau|r_{i})=e^{-\lambda_{\text{within}}\tau}$ when
$0<\tau<1$ and $\tau_{ij}=\infty$ with probability $e^{-\lambda_{\text{within}%
}}$. \ When $i$ and $j$ are in different clusters, $\tau_{ij}$ has an
exponential distribution with hazard $\lambda_{\text{between}}$ truncated at
$r_{i}$, so $S_{ij}(\tau|r_{i})=e^{-\lambda_{\text{between}}\tau}$ when
$0<\tau<1$ and $\tau_{ij}=\infty$ with probability $e^{-\lambda
_{\text{between}}}$. \ 

We fixed the hazard of infectious contact between individuals in the same
cluster at $\lambda_{\text{within}}=.4$. \ We tuned the hazard of infectious
contact between individuals in different clusters to obtain $R$ mean
infectious contacts by infectious individuals; specifically,
\[
\lambda_{\text{between}}=\frac{R-(k-1)(1-e^{-.4})}{n-k}.
\]
We chose $\lambda_{\text{within}}=.4$ to obtain rapid transmission within
clusters while retaining sufficient transmission between clusters to sustain
an epidemic. \ Note that when $k>R(1-e^{-.4})^{-1}+1$, we get the implausible
result that $\lambda_{\text{between}}<0$. \ Clearly, $R$ and $k$ must be
chosen so that an infectious person makes an average of $R$ or fewer
infectious contacts within his or her cluster, which guarantees that
$\lambda_{\text{between}}\geq0$.

At a given $R$, the mean infectious contact interval given that infectious
contact occurs depends on the cluster size. \ If the entire population is
infectious and the cluster size is $k$, then a given individual will receive
an average of $R$ infectious contacts, of which $(k-1)(1-e^{-.4})$ come from
within his or her cluster. \ The mean infectious contact interval for
within-cluster contacts is
\[
\frac{1}{1-e^{-.4}}\int_{0}^{1}.4\tau e^{-.4\tau}d\tau,
\]
and the mean infectious contact interval for between-cluster contacts is
approximately $.5$ (as in the models for global competition). \ Therefore, the
mean infectious contact interval given that contact occurs and the cluster
size is $k$ is
\[
E[\tau_{ij}|\tau_{ij}<\infty,k]\approx(1-\frac{(k-1)(1-e^{-.4})}{R}%
).5+\frac{(k-1)}{R}\int_{0}^{1}.4\tau e^{-.4\tau}d\tau.
\]
To compare generation interval contraction for different cluster sizes, we
calculated \textit{scaled generation intervals} by dividing the observed
generation intervals at each cluster size by $E[\tau_{ij}|\tau_{ij}<\infty
,k]$. \ If the mean generation interval remained constant, we would expect the
mean scaled generation interval to be approximately one throughout an epidemic.

\begin{figure}
[ptb]
\begin{center}
\includegraphics[
height=3.7602in,
width=5.5478in
]%
{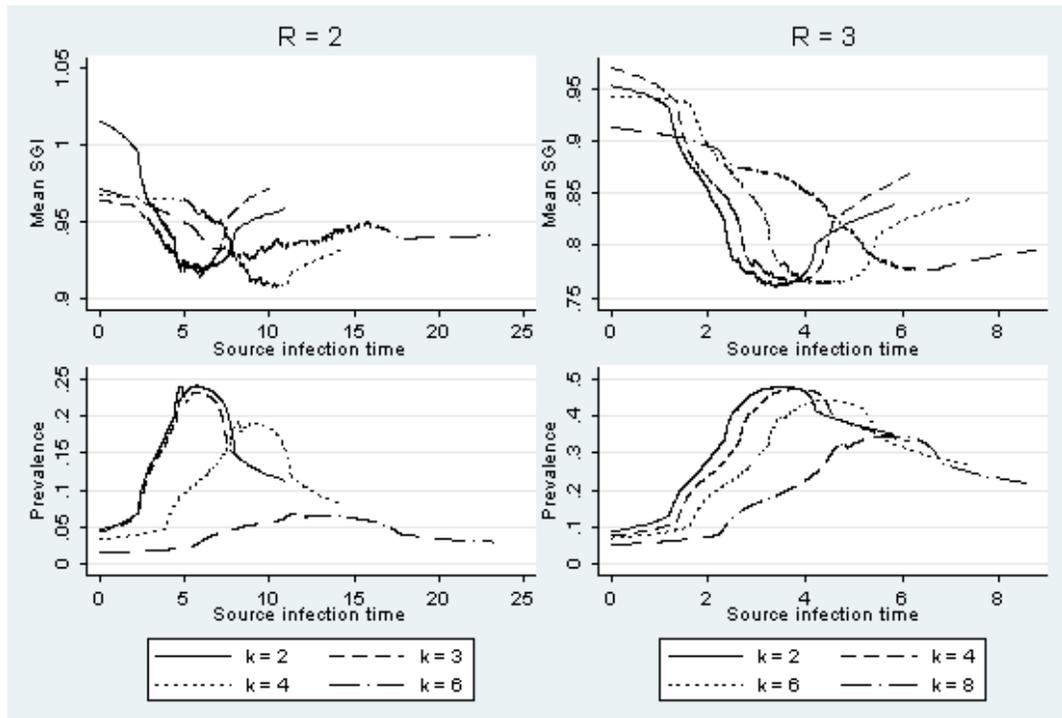}%
\caption{The smoothed mean scaled generation interval (SGI) and prevalence as
a function of the source infection time for $R=2$ and $R=3$. \ With increasing
cluster size, the degree of generation interval contraction is roughly the
same even though the peak prevalence of infection is lower. }%
\label{local}%
\end{center}
\end{figure}

For $R=2$, we ran the model with cluster sizes of $1$ through $6$. \ For $R=3
$, we ran the model with cluster sizes of $2$ through $8$. \ For each
simulation, we recorded $t_{i}$, $v_{i}$, $t_{v_{i}}$, and the prevalence of
infection at time $t_{i}$ in each infector/infectee pair. \ Figure \ref{local}
shows smoothed mean curves for the generation interval and prevalence versus
the source infection time for several cluster sizes at each $R$. \ As before,
there is a clear tendency of the mean generation interval to contract. \ The
degree of contraction is roughly the same for all cluster sizes, but this
contraction is maintained at a lower global prevalence of infection in models
with larger cluster sizes. \ Similar results were obtained for cluster sizes
not shown. \ Again, these results are exactly in line with the argument of
Section 3.

\section{Consequences for estimation}

The effect of generation interval contraction on parameter estimates obtained
from models that assume a constant generation or serial interval distribution
is difficult to assess. \ The assumption of a constant serial or generation
interval distribution may be reasonable in the early stages of an epidemic
with little clustering of contacts, in an epidemic with $R_{0}$ near one, or
in an endemic situation. \ However, this ignores the more fundamental issue
that estimates of these distributions are obtained from transmission events
where the infector/infectee pairs are known (often because of transmission
from a known patient within a household or hospital ward). \ Even in the early
stages of an epidemic, the generation interval distribution in these settings
may differ substantially from the generation interval distribution for
transmission in the general population.

In this section, we argue that hazards of infectious contact can be used
instead of generation or serial intervals in the analysis of epidemic data.
\ As an example, we look at the estimator of $R(t)$ (the effective
reproductive number at time $t$) derived by Wallinga and Teunis
\cite{Wallinga} and applied to data on the SARS outbreaks in Hong Kong,
Vietnam, Singapore, and Canada in 2003. \ In their paper, the available data
was the \textquotedblleft epidemic curve\textquotedblright\ $\mathbf{t}%
=(t_{(1)},...,t_{(m)})$, where $t_{(i)}$ is the infection time of the
$i^{\text{th}}$ person infected. \ They assume a probability density function
(pdf) $w(\tau|\mathbf{\theta})$ for the serial interval given a vector
$\mathbf{\theta}$ of parameters (note that this parameter vector applies to
the population, not to individuals). \ The infector of person $(i)$ is denoted
by $v_{(i)}$, with $v_{(i)}=0$ for imported infections. \ The
\textquotedblleft infection network\textquotedblright\ is a vector
$\mathbf{v}=(v_{(1)},...,v_{(m)})$ specifying the source of infection for each
infected person. \ With these assumptions, the likelihood of $\mathbf{v}$ and
$\mathbf{\theta}$ given $\mathbf{t}$ is
\[
L(\mathbf{v},\mathbf{\theta}|\mathbf{t})=\prod_{i:v_{(i)}\neq0}w(t_{(i)}%
-t_{v_{(i)}}|\mathbf{\theta}).
\]
The sum of this likelihood over the set $V$ of all infection networks
consistent with the epidemic curve $\mathbf{t}$ is%
\[
L(\mathbf{\theta}|\mathbf{t})=\prod_{i:v_{(i)}\neq0}\sum_{j\neq i}%
w(t_{i}-t_{j}|\mathbf{\theta}).
\]
Taking a likelihood ratio, Wallinga and Teunis argue that the relative
likelihood that person $k$ was infected by person $j$ is
\begin{equation}
p_{jk}^{(WT)}=\frac{w(t_{k}-t_{j}|\mathbf{\theta})}{\sum_{i\neq k}%
w(t_{k}-t_{i}|\mathbf{\theta})}.\label{pjkWallinga}%
\end{equation}
The number $R_{j}$ of secondary infectious generated by person $j$ is a sum of
Bernoulli random variables with expectation
\[
E[R_{j}]=\sum_{k=1}^{n}p_{jk}^{(WT)}.
\]
An estimate of the effective reproductive number $R(t)$ can be obtained by
calculating a smoothed mean for a scatterplot of $(t_{j},E[R_{j}])$. \ This
analysis is ingenious, but it can be only approximately correct because the
distribution of serial intervals varies systematically over the course of an epidemic.

\subsection{Hazard-based estimator}

A very similar result can be derived by applying the theory of order
statistics (see Ref. \cite{Gut}) to the general stochastic SIR model from
Section 2. \ Specifically, we use the following results: If $X_{1},...,X_{n}$
are independent non-negative random variables, then their minimum $X_{(1)}$
has the hazard function
\[
\lambda_{(1)}(t)=\sum_{i=1}^{n}\lambda_{i}(t).
\]
Given that the minimum is $x_{(1)}$, the probability that $X_{j}=x_{(1)}$
(i.e. that the minimum was observed in the $j^{\text{th}}$ random variable)
is
\[
\frac{\lambda_{j}(x_{(1)})}{\sum_{i=1}^{n}\lambda_{i}(x_{(1)})}.
\]
For simplicity, we assume that the infectious contact intervals $\tau_{ij}$
are absolutely continuous random variables. \ 

Let $\lambda_{ij}(\tau|r_{i})$ be the conditional hazard function for
$\tau_{ij}$ given $r_{i}$ and let $\lambda_{0i}(t)$ be the hazard function for
infectious contact to $i$ from outside the population at time $t$. \ Since
$\tau_{ij}$ is nonnegative, $\lambda_{ij}(\tau|r_{i})=0$ whenever $\tau<0$.
\ Let $H(t)$ denote the set of infection times and recovery periods for all
$i$ such that $t_{i}\leq t$. \ If person $k$ is susceptible at time $t$, his
or her total hazard of infection at time $t$ given $H(t)$ is $\sum_{i=0}%
^{n}\lambda_{ik}(t-t_{i}|r_{i})$, where we let $\lambda_{0k}(t-t_{0}%
|r_{0})=\lambda_{0k}(t)$ for simplicity of notation. \ If an infection occurs
in person $k$ at time $t_{k}<\infty$, then the conditional probability that
person $j$ infected person $k$ given $H(t_{k})$ is
\begin{equation}
p_{jk}=\frac{\lambda_{jk}(t_{k}-t_{j}|r_{j})}{\sum_{i=0}^{n}\lambda_{ik}%
(t_{k}-t_{i}|r_{i})},\label{pjk}%
\end{equation}
which is the probability that $t_{jk}=\min(t_{0k},t_{1k},...,t_{nk})$. \ This
has the same form as equation (\ref{pjkWallinga}) except that it uses hazards
of infectious contact instead of a pdf for the serial interval. \ If the
hazards of infectious contact in the underlying SIR model do not change over
the course of an epidemic, then $p_{jk}$ can be estimated accurately
throughout an epidemic. \ Unlike the assumption of a stable generation or
serial interval distribution, this assumption is unaffected by competition
among potential infectors. \ The rest of the estimation of $R(t)$ could
proceed exactly as in Ref. \cite{Wallinga}, replacing $p_{jk}^{(WT)}$ with
$p_{jk}$. \ 

\subsection{Partial likelihood for epidemic data}

A partial likelihood for epidemic data can be derived using the same logic as
that used to derive $p_{jk}$ in equation (\ref{pjk}). \ For each person $k $
such that $t_{k}<\infty$, the probability that the failure at time $t_{k}$
occurred in person $k$ given $H(t_{k})$ is
\begin{equation}
\frac{\sum_{i=0}^{n}\lambda_{ik}(t_{k}-t_{i}|r_{i})}{\sum_{j=1}^{n}\sum
_{i=0}^{n}\lambda_{ij}(t_{k}-t_{i}|r_{i})},\label{Pfail}%
\end{equation}
where the numerator is the hazard of infection (from all sources) in person
$k$ at time $t_{k}$ and the denominator is the total hazard of infection for
all persons at risk of infection at time $t_{k}$. \ 

If there is a vector of parameters $\mathbf{x}_{ij}$ for each pair $ij$ (which
may include individual-level covariates for $i$ and $j$ as well as pairwise
covariates for the ordered pair $ij$) and a vector of parameters
$\mathbf{\theta}$ such that $\lambda_{ij}(\tau|r_{i})=\lambda(\tau
|r_{i},\mathbf{x}_{ij},\mathbf{\theta})$, then a partial likelihood for
$\mathbf{\theta}$ can be obtained by multiplying equation (\ref{Pfail}) over
all $m$ observed failure times. \ If $(k)$ denotes the index of the
$k^{\text{th}}$ person infected, $\mathbf{t}=(t_{1},...,t_{n})$, and
$\mathbf{X}=\{\mathbf{x}_{ij}:i,j=1,...,n\}$, then the partial likelihood is
\begin{equation}
L_{p}(\mathbf{\theta}|\mathbf{t},\mathbf{X})=\prod_{k=1}^{m}\frac{\sum
_{i=0}^{n}\lambda(t_{(k)}-t_{i}|r_{i},\mathbf{x}_{i(k)},\mathbf{\theta})}%
{\sum_{j=1}^{n}\sum_{i=0}^{n}\lambda(t_{k}-t_{i}|r_{i},\mathbf{x}%
_{ij},\mathbf{\theta})}.\label{Lp}%
\end{equation}
This is very similar to partial likelihoods that arise in survival analysis,
so many techniques from survival analysis may be adaptable for use in the
analysis of epidemic data. \ 

The goal of such methods would be to allow statistical inference about the
effects of individual and pairwise covariates on the hazard of infection in
ordered pairs of individuals. \ In the ordered pair $ij$, the effects of
individual covariates for $i$ and $j$ on $\lambda_{ij}(\tau|r_{i})$ would
reflect the infectiousness of $i$ and the susceptibility of $j$, respectively.
\ Pairwise covariates could include such information as whether $i$ and $j$
are in the same household, the distance between their households, whether they
are sexual partners, and any other aspects of their relationship to each other
thay may affect the hazard of infection from $i$ to $j$. \ 

This approach has several advantages over any approach based on a distribution
of generation or serial intervals. \ First, it is not necessary to determine
who infected whom in any subset of observed infections. \ If $v_{j}$ is known
for some $j$, this knowledge can be incorporated in the partial likelihood by
replacing the term for the failure time of person $j$ in (\ref{Lp}) with
$p_{v_{j}j}$ from equation (\ref{pjk}). \ Second, this approach allows the use
individual-level and pairwise covariates for inference in a flexible and
intuitive way. \ The resulting estimated hazard functions have a
straightforward interpretation and can be incorporated naturally into a
stochastic SIR model. \ Third, this approach allows theory and methods from
survival analysis to be applied to the analysis of epidemic data. \ 

\section{Discussion}

Generation and serial interval distributions are not stable characteristics of
an infectious disease. \ When multiple infectious persons compete to infect a
given susceptible person, infection is caused by the first person to make
infectious contact. \ In Section 3, we showed that the mean infectious contact
interval $\tau_{ij}$ given that $i$ actually infected $j$ is less than or
equal to the mean $\tau_{ij}$ given $i$ made infectious contact with $j$.
\ That is,
\[
E[\tau_{ij}|v_{j}=i]\leq E[\tau_{ij}|\tau_{ij}<\infty],
\]
with strict inequality when $\tau_{ij}$ is non-constant and $j$ is at risk of
infectious contact from any source other than $i$ (more precise conditions are
given in Section 3). \ This result holds for all time-homogeneous stochastic
SIR models. \ 

In an epidemic, the mean generation (and serial) intervals contract as the
prevalence of infection increases and susceptible persons are at risk of
infectious contact from multiple sources. \ In the simulations of Section 4,
we saw that the degree of contraction increases with $R_{0}$. \ For models
with clustering of contacts, generation interval contraction can occur even
when the global prevalence of infection is low because susceptibles are at
risk of infectious contact from multiple sources within their own clusters.
\ In all of the simulations, the greatest serial interval contraction
coincided with the peak prevalence of infection, when the risk of infectious
contacts from multiple sources was highest. \ The mean generation interval
increases again as the epidemic wanes, but this rebound may be small when
$R_{0}$ is high. \ 

The reason that generation and serial intervals contract during an epidemic is
that their definition applies to pairs of individuals $ij$ such that $i$
actually transmitted infection to $j$. \ If we don't require that an
infectious contact leads to the transmission of infection, we are led
naturally to the concept of the infectious contact interval, which has a
well-defined distribution throughout an epidemic. \ Similarly, we can define
$R_{0}$ as the mean number of infectious contacts (i.e., finite infectious
contact intervals) made by a primary case without reference to a completely
susceptible population. \ Generation and serial intervals and the effective
reproductive number can then be defined in terms of infectious contacts that
actually lead to the transmission of infection. \ Many fundamental concepts in
infectious disease epidemiology can be simplified usefully by defining them in
terms of infectious contact rather than infection transmission.

Infectious contact hazards for ordered pairs of individuals can be used for
many of the same types of analysis that have been attempted using generation
or serial interval distributions. \ In Section 5, We derived a hazard-based
estimator of $R(t)$ very similar to that developed by Wallinga and Teunis
\cite{Wallinga}. \ This derivation led naturally to a partial likelihood for
epidemic data very similar to those that arise in survival analysis. \ We
believe that the adaptation of methods and theory from survival analysis to
infectious disease epidemiology will yield flexible and powerful tools for
epidemic data analysis.

\textbf{Acknowledgements:} \textit{This work was supported by the US National
Institutes of Health cooperative agreement 5U01GM076497 "Models of Infectious
Disease Agent Study" (E.K. and M.L.) and Ruth L. Kirchstein National Research
Service Award 5T32AI007535 "Epidemiology of Infectious Diseases and
Biodefense" (E.K.).} \ \textit{We also wish to thank Jacco Wallinga and the
anonymous reviewers of Mathematical Biosciences for useful comments and
suggestions.}

\appendix{}\ 

\section{Figures and tables}%

\begin{table}[p] \centering
\begin{tabular}
[c]{|l|l|l|}\hline
$\mathbf{R}_{0}$ & $\mathbf{E[\tau}_{ij}|\mathbf{\tau}_{ij}<\infty\mathbf{]}$
& $.5-\mathbf{E[\tau}_{ij}|\mathbf{\tau}_{ij}<\infty\mathbf{]}$\\\hline
$1.25$ & $.49999$ & $.00001$\\\hline
$1.5$ & $.499988$ & $.000012$\\\hline
$2$ & $.499983$ & $.000017$\\\hline
$3$ & $.499975$ & $.000025$\\\hline
$4$ & $.499967$ & $.000033$\\\hline
$5$ & $.499958$ & $.000042$\\\hline
\end{tabular}
\caption{Expected infectious contact interval given that infectious contact occurs
in the models illustrating global competition among potential infectors.
If the generation interval were constant, this would be the mean generation interval throughout an epidemic.}\label{Expected}%
\end{table}%
%

%

%

%

%

\end{document}